# High Thermal Conductivity of Rutile-GeO$_2$ Film by MOCVD: 52.9 W m$^{-1}$ K$^{-1}$

Imteaz Rahaman[1,*], Michael E. Liao[2,3,*], Ziqi Wang[4,*], Eugene Y. Kwon[3], Rui Sun[5], Botong Li[1], Hunter D. Ellis[1], Bobby G. Duersch[6], Dali Sun[5], Jun Liu[4], Mark S. Goorsky[3], Michael A. Scarpulla[1,7], and Kai Fu[1, a)]

[1]Department of Electrical and Computer Engineering, The University of Utah, Salt Lake City, UT 84112, USA

[2]Apex Microdevices, West Chester, Ohio, 45069, USA

[3]Department of Materials Science and Engineering, University of California, Los Angeles, California, 90095, USA

[4]Department of Mechanical and Aerospace Engineering, North Carolina State University, Raleigh, NC 27695, USA

[5]Department of Physics, North Carolina State University, Raleigh, NC 27695, USA

[6]Electron Microscopy and Surface Analysis Laboratory, The University of Utah, Salt Lake City, UT 84112, USA

[7]Department of Materials Science and Engineering, The University of Utah, Salt Lake City, UT 84112, USA

**Abstract**

Rutile germanium dioxide (r-GeO$_2$) has recently emerged as a promising ultrawide-bandgap (UWBG) semiconductor owing to its wide bandgap (~4.4–5.1 eV), ambipolar doping potential, and high theoretical thermal conductivity. However, experimental data on the thermal conductivity of r-GeO$_2$ epitaxial layers have not been reported, primarily due to challenges in phase control and surface roughness. Here, we report the high thermal conductivity of 52.9 ± 6.6 W m$^{-1}$ K$^{-1}$ for high-quality (002) r-GeO$_2$ films grown by metal–organic chemical vapor deposition (MOCVD) and characterized using time-domain thermoreflectance (TDTR). The phase control was achieved through a seed-driven stepwise crystallization (SDSC) approach, and the surface roughness was significantly reduced from 76 nm to 16 nm (locally as low as 1 Å) via chemical mechanical polishing (CMP). These results highlight the promise of r-GeO$_2$ as a UWBG oxide platform for power electronics applications.

* These authors contributed equally to this work.

a) Author to whom correspondence should be addressed. Electronic mail: kai.fu@utah.edu

Ultrawide bandgap (UWBG) semiconductors have drawn substantial attention for next-generation power and high-frequency electronics due to their large breakdown fields, high carrier mobilities, and stability under extreme conditions. Materials such as GaN, AlN, β-Ga$_2$O$_3$, diamond, and c-BN have shown remarkable performance;[1–5] however, each presents intrinsic challenges that limit further development. GaN and AlGaN face thermal-management challenges during high-power operation.[6,7] β-Ga$_2$O$_3$ suffers from both poor p-type conductivity and low thermal conductivity,[8–12] while diamond, though highly thermally conductive, remains difficult to process and integrate into device structures[13,14]. These limitations have motivated exploration of alternative UWBG oxides that combine a wide bandgap with both carrier polarity and improved heat dissipation. Among them, rutile germanium dioxide (r-GeO$_2$) has recently emerged as a promising candidate. It offers a bandgap of approximately 4.4–5.1 eV[15–18], high calculated electron mobilities of 244 cm²/V·s (⊥c) and 377 cm²/V·s (∥c),[19] and the potential for ambipolar doping through relatively shallow donor and acceptor ionization energies. Furthermore, its predicted Baliga figure of merit (~1.4 × 10² GW cm$^{-2}$)[20] exceeds that of β-Ga$_2$O$_3$, indicating superior intrinsic potential for power-device applications. In addition, its structural and chemical compatibility with other rutile oxides such as TiO$_2$ and SnO$_2$,[21–25] enables high-quality heterojunctions, positioning r-GeO$_2$ as a compelling platform for advanced power electronics..

Epitaxial growth of r-GeO$_2$ thin films has been explored by several techniques, including molecular beam epitaxy (MBE)[26], pulsed laser deposition (PLD)[27,28], metal-organic chemical vapor deposition (MOCVD)[29,30], and mist chemical vapor deposition (Mist-CVD)[31,32]. Despite these efforts, achieving single-phase, fully crystalline rutile films remains challenging due to the small thermodynamic energy separation between the quartz, amorphous, and rutile polymorphs, as well as differences in their surface energies.[33] Consequently, many studies report partial phase conversion or mixed crystallinity. Theoretical work predicts that r-GeO$_2$ possesses anisotropic phonon-limited thermal conductivities of 37 W m$^{-1}$ K$^{-1}$ along the *a*-axis (in-plane) and 58 W m$^{-1}$ K$^{-1}$ along the *c*-axis (out-of-plane),[34] with a directionally averaged value near 44 W m$^{-1}$ K$^{-1}$. Experimental measurements on bulk polycrystalline pellets yielded approximately 51 W m$^{-1}$ K$^{-1}$ at 300 K, in close agreement with the theoretical average.[34] However, the thermal properties of epitaxial r-GeO$_2$ films have not yet been reported, primarily due to issues with phase control and surface roughness.

In this study, we employed a seed-driven stepwise crystallization (SDSC) approach [29,33] to stabilize the rutile phase of GeO$_2$ and chemical mechanical polishing (CMP) to reduce surface roughness. The structural properties were evaluated using high-resolution X-ray diffraction (XRD), rocking curve (ROC), reciprocal-space mapping (RSM), phi-scan, and cathodoluminescence (CL) to examine crystalline alignment, strain relaxation, and defect distribution. The thermal conductivity was measured by the time-domain thermoreflectance (TDTR) method.

Rutile GeO$_2$ thin films were prepared by metal–organic chemical vapor deposition (MOCVD) using an Agnitron Agilis system. The films were grown using a previously established seed-driven stepwise crystallization (SDSC) strategy, which enabled uniform rutile-phase formation across the substrate. All growth segments were carried out at 925 °C under a constant chamber pressure of 80 Torr. Tetraethylgermane (TEGe) and high-purity $O_2$ were used as the Ge and oxidant precursors, respectively, with argon (Ar) serving as both the carrier and shroud gas. The precursor flow rates were maintained at $1.35 \times 10^{-5}$ mol min$^{-1}$ for TEGe and $8.94 \times 10^{-2}$ mol min$^{-1}$ for $O_2$, and the susceptor rotation speed was fixed at 300 rpm to ensure uniform deposition across the substrate surface.

Chemical mechanical polishing (CMP) of $GeO_2$ films was performed using a NaOH-based colloidal silica slurry on a Logitech PM5 CMP tool. The slurry flow rate was 10 mL/min with a pad rotation speed of 30 RPM using a polyurethane polishing pad. The applied downward force was ~5 kPa. Polishing parameters employed were similar to those used in previous work on the CMP optimization of β-$Ga_2O_3$ published earlier[35].

The thermal conductivity of rutile $GeO_2$ films was measured using the TDTR method. Details of this technique and the experimental setup can be found in previous works[36–38]. Prior to measurement, a ~80 nm-thick Al thin film was thermally evaporated onto the samples at a rate of 0.2 nm/s with a base chamber pressure better than $5 \times 10^{-8}$ mTorr. A mode-locked Ti:sapphire laser generates a train of pulses at a repetition rate of 80 MHz. The laser is split into pump and probe beams, and a mechanical delay stage is used to change the optical path difference between the two beams before being focused onto the sample surface through the 5× objective lens (both beams have a Gaussian beam spot size of ≈12 μm after focusing). The probe beam detects the periodic reflectivity change of the Al transducer from the pump at a modulation frequency ($f$ = 7.1 MHz) as a function of the delay time. The 7.1 MHz modulation frequency is chosen based on the sensitivity analysis to reduce the sensitivity of substrate thermal properties to the measured signal.

The time decay of the probed signal is used to determine thermal conductivity by fitting it with a multilayered heat conduction model. The thickness of the Al transducer is determined using picosecond acoustics. The thermal properties of the Al and substrate (r-TiO$_2$) are adopted from the literature values.[39] The reported total uncertainty of the measured thermal conductivity is calculated by adding the measurement uncertainties (among multiple spots and scans) in quadrature with the systematic errors that propagate from uncertainties in the geometrical and thermophysical parameters, such as metal film thickness, laser spot size, and thermal properties of the metal transducer and substrate.

Figure 1 summarizes the surface morphology evolution of the rutile GeO$_2$ film before and after the chemical mechanical polishing (CMP) process. The as-grown film (~2 μm thick) deposited on r-TiO$_2$ (001) exhibits faceted grain (faceted crystal shape) morphology (Fig. 1a-b) with a high surface roughness of RMS ≈ 76 nm (Fig. 1b). Such roughness originates from non-uniform lateral growth and local island coalescence inherent to vapor-phase deposition at elevated temperatures. These surface features promote phonon scattering at grain boundaries and interfaces, which can significantly limit accurate thermal-conductivity measurements. After CMP, the film thickness is reduced to approximately 1.1 μm, and the surface becomes notably smoother (Fig. 1c). The intermediate polishing stage (Fig. 1d) shows RMS ≈ 16 nm, indicating the progressive removal of large asperities. In addition, the surface achieves an atomic-scale roughness of ~1 Å locally (Fig. 1e), revealing a mirror-like finish suitable for optical-pump measurements such as TDTR. The improvement in surface morphology reflects the combined effect of mechanical abrasion and chemical etching during CMP, which removes high-energy grain peaks while maintaining the continuity of the rutile phase. Such planarization minimizes optical interference and ensures consistent heat flow across the film during transient thermal measurements.

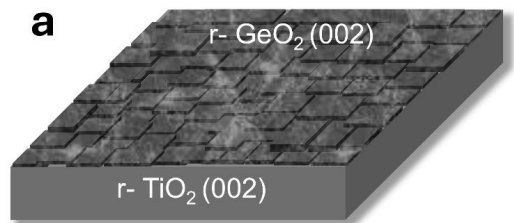

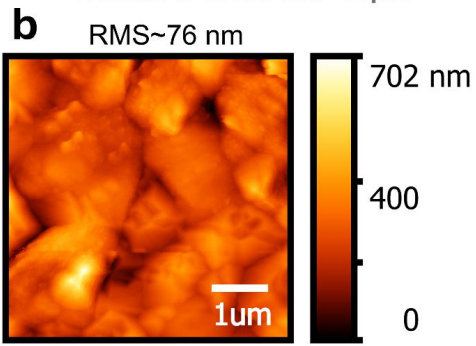

**CMP**

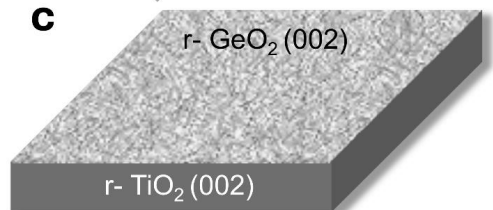

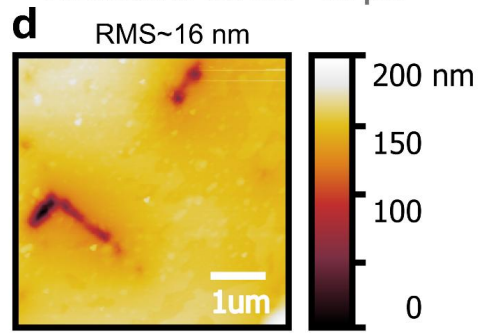

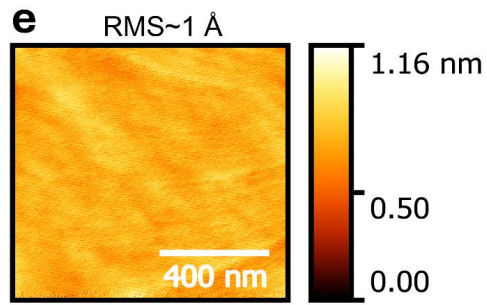

**FIG. 1.** Surface morphology evolution of rutile GeO$_2$ film before and after CMP. (a) Schematic of as-grown r-GeO$_2$ (002) film on r-TiO$_2$ (002) substrate (thickness ≈ 2 µm). (b) AFM image of the as-grown surface showing faceted morphology with RMS ≈ 76 nm. (c) Schematic of the polished r-GeO$_2$ film (thickness ≈ 1.1 µm). (d) Intermediate CMP stage exhibiting reduced roughness (RMS ≈ 16 nm). (e) Polished surface exhibiting locally smooth regions approaching atomic-scale roughness (RMS ≈ 1 Å). The CMP process effectively removes surface irregularities and produces a mirror-flat surface suitable for high-precision TDTR thermal characterization.

Figure 2 presents the crystallographic analysis of rutile GeO$_2$ films before and after the chemical mechanical polishing (CMP) process. The *2θ-ω* scan in Fig. 2a confirms that both samples exhibit dominant r-GeO$_2$ (002) and r-TiO$_2$ (002) reflections without secondary impurity peaks, indicating that the rutile phase remains stable after CMP. Minor quartz-GeO$_2$ peaks observed in the as-grown film are no longer detectable after polishing, indicating that the CMP process effectively dissolves the water-soluble quartz phase and removes weakly adhered surface grains during planarization. The rocking-curve (ω-scan) of the r-GeO$_2$ (002) reflection (Figure 2b) reveals that both the as-grown and CMP-treated films exhibit comparable full width at half maximum (FWHM) values of approximately 0.28°, indicating that the chemical-mechanical polishing process did not alter the crystalline quality of the r-GeO$_2$ layer. This result confirms that surface planarization effectively thinned and smoothed the film surface without introducing additional lattice distortion or dislocation broadening. The preservation of a narrow FWHM after CMP further suggests that the underlying crystal alignment and epitaxial registry with the r-TiO$_2$ (002) substrate remain intact. The φ-scan of the (202) reflection (Figs. 2c, d) retains a fourfold symmetry characteristic of the rutile structure, confirming that the in-plane epitaxial relationship between GeO$_2$ and TiO$_2$ remains unchanged after CMP.

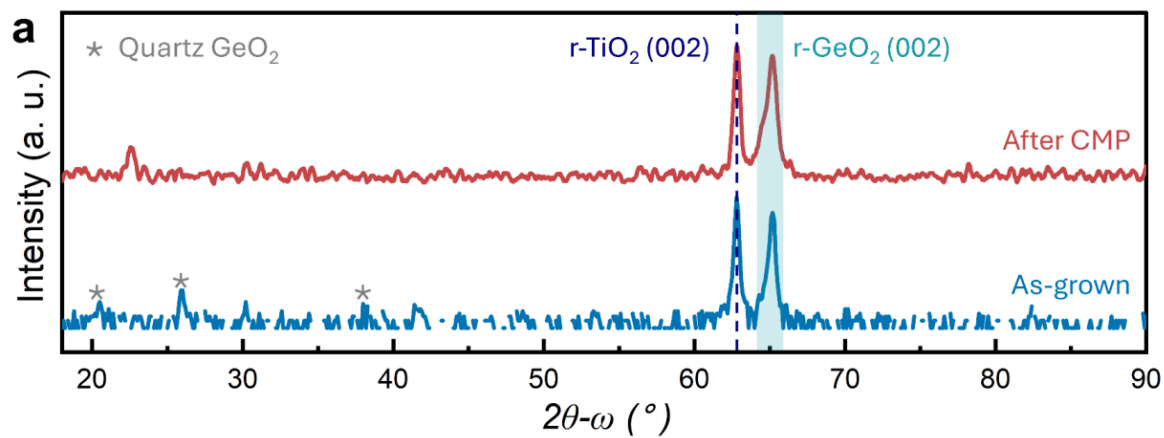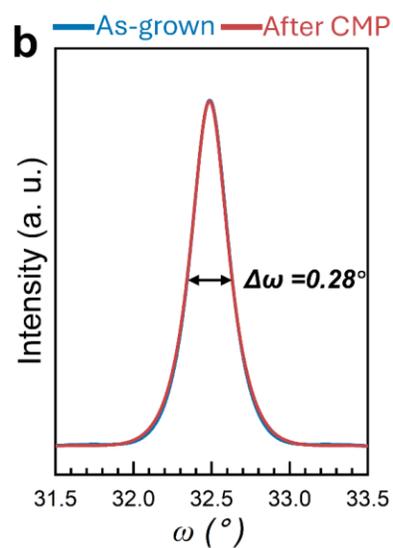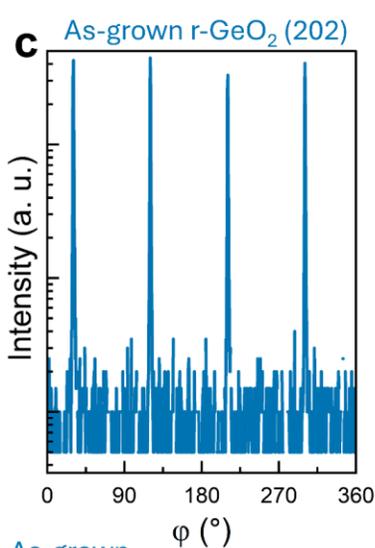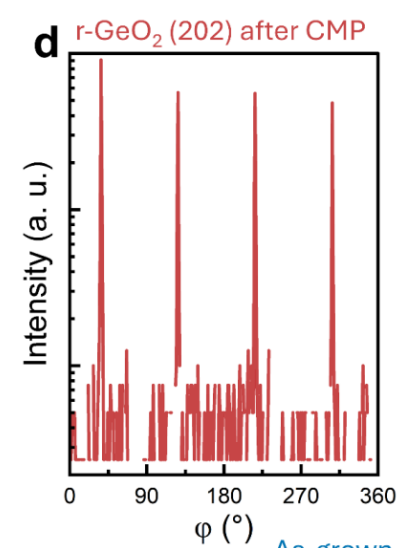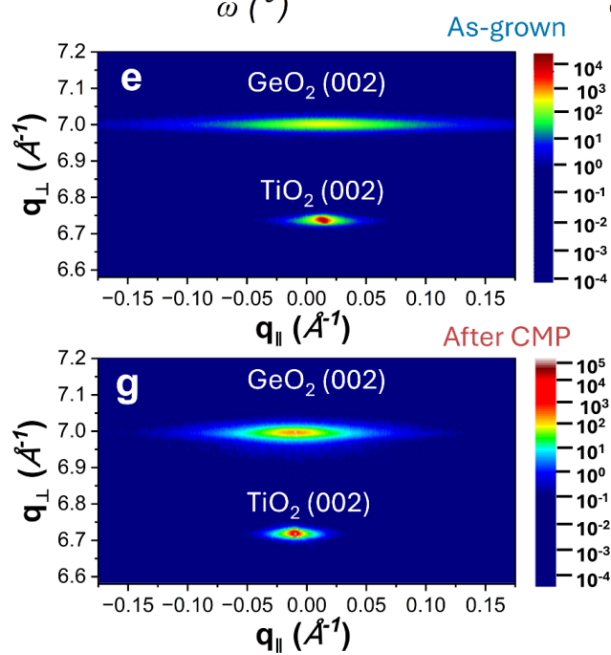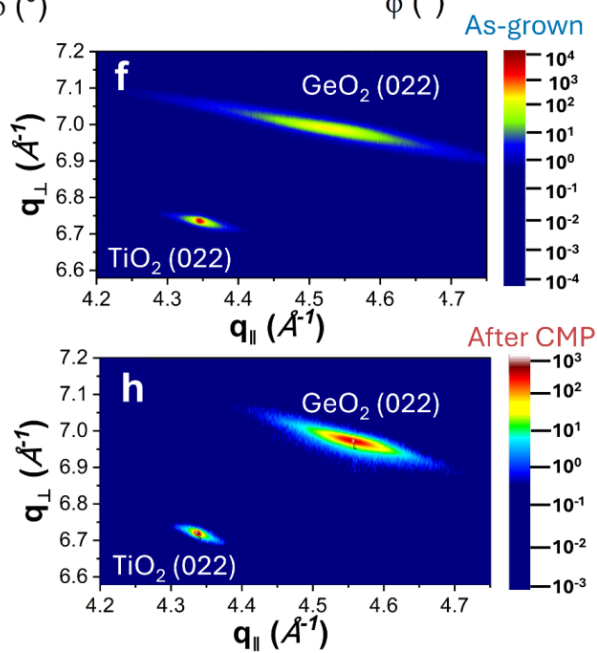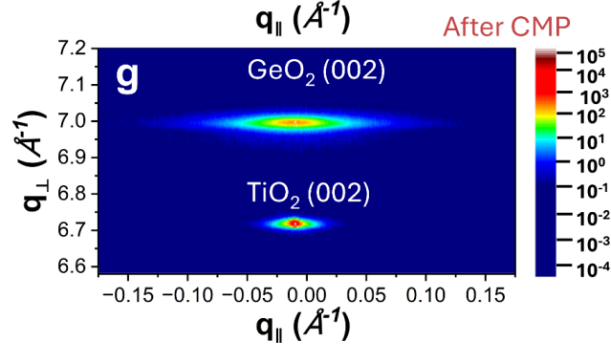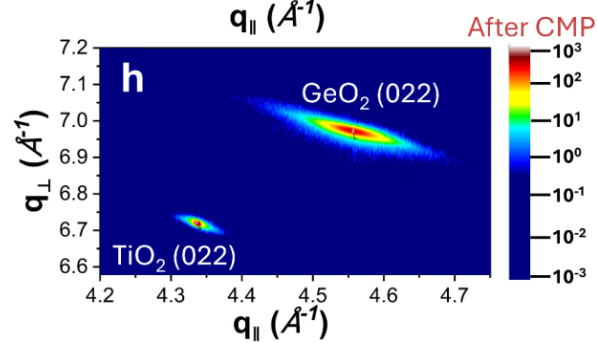

**FIG. 2.** Structural characterization of rutile GeO$_2$ films before and after CMP. (a) $2\theta$-$\omega$ scans showing r-GeO$_2$ (002) and r-TiO$_2$ (002) reflections; minor quartz peaks in the as-grown film disappear after CMP. (b) Rocking-curve comparison of r-GeO$_2$ (002) showing similar FWHM (~0.28°) for as-grown and after CMP, confirming unchanged crystalline quality. (c,d) φ-scans of r-GeO$_2$ (202) reflection before and after CMP, confirming fourfold symmetry and epitaxial alignment. (e–h) Reciprocal-space maps of r-GeO$_2$ and r-TiO$_2$ for the symmetric (002) and asymmetric (022) reflections of the as-grown (e, f) and CMP-treated (g, h) films, showing coherent epitaxy and partial relaxation after polishing.

Reciprocal-space mapping (RSM) of the symmetric (002) and asymmetric (022) reflections (Figs. 2e–h) further reveals that the r-GeO$_2$ peaks stay aligned with those of r-TiO$_2$, signifying coherent epitaxy with partial relaxation. The peak intensity enhancement and sharpening observed after polishing suggest improved surface uniformity and reduced diffuse scattering.

To understand the optical response and defect-related emission in r-GeO$_2$ films, cathodoluminescence (CL) spectra were analyzed before and after chemical mechanical polishing (CMP), as summarized in Figure 3. The CL characteristics of the as-grown film are discussed in our previous publication[40]. For reference, the CL spectrum of the underlying r-TiO$_2$ (001) substrate exhibits three main emission bands (D$_{T1}$–D$_{T3}$) centered at approximately 2.80 eV, 2.56 eV, and 2.10 eV, respectively (from prior measurement)[40]. These features correspond to oxygen-vacancy–related deep-level transitions and trap-assisted recombination pathways characteristic of rutile TiO$_2$, consistent with earlier reports [41,42]. The polished r-GeO$_2$ film (Fig. 3a) displays two dominant blue–green emissions, labeled D$_{G1}$ (~2.64 eV, 470 nm) and D$_{G2}$ (~2.38 eV, 520 nm), [43,44]. These emissions are typically associated with intrinsic or defect-related luminescence centers within r-GeO$_2$ and have been previously observed under both cathodoluminescence and photoluminescence excitation.

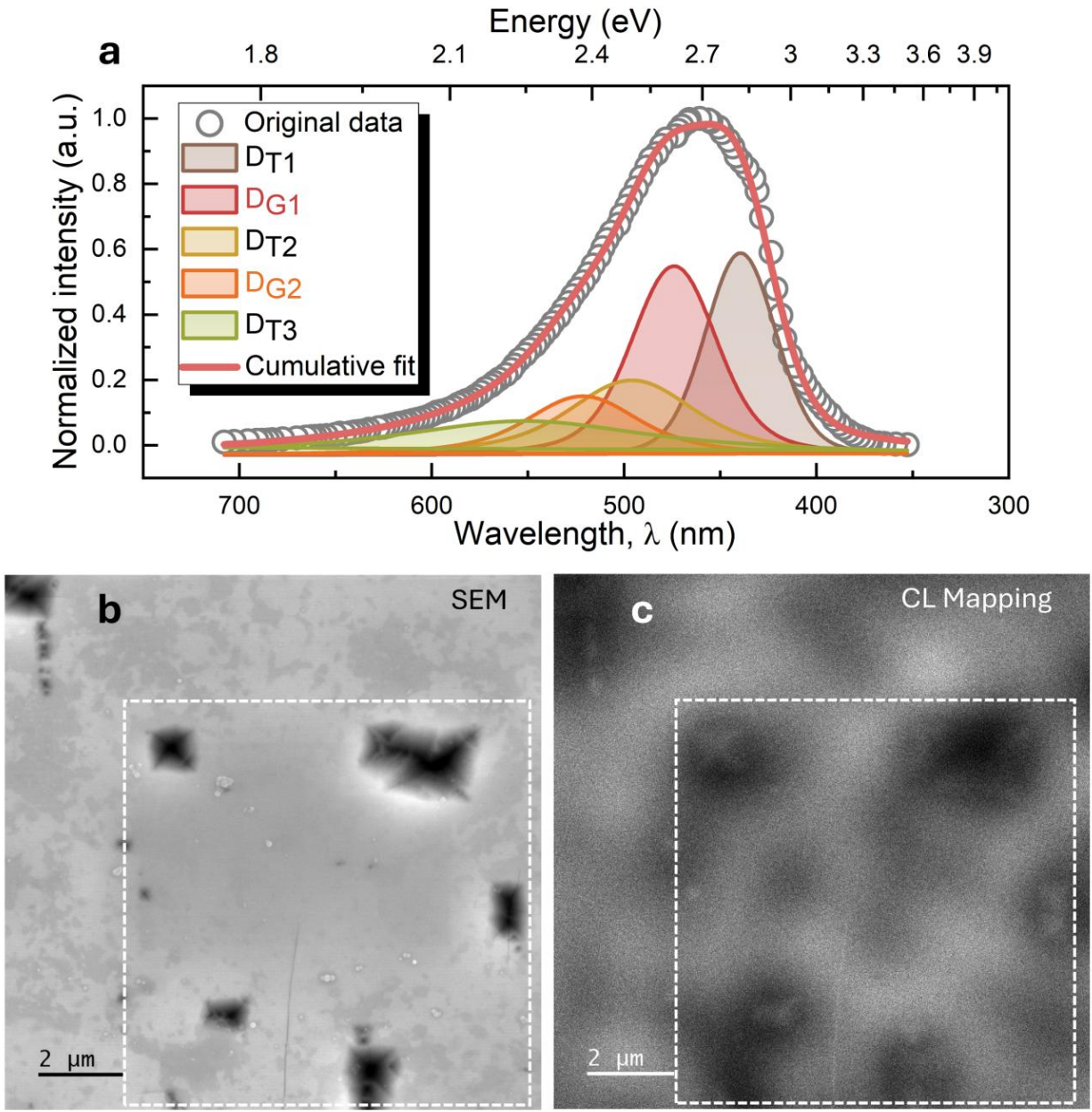

**FIG. 3.** Cathodoluminescence and surface morphology of r-GeO$_2$ after CMP. (a) Normalized CL spectrum of the polished film deconvoluted into Gaussian components where DT$_1$–DT$_3$ originate from the r-TiO$_2$ substrate and DG$_1$–DG$_2$ from the r-GeO$_2$ films, showing broad blue–green emissions near 470 and 520 nm. (b) SEM image of the CMP-treated surface displaying smooth morphology with sparse nanoscale pits. (c) Corresponding CL intensity map revealing uniform emission and improved optical homogeneity after polishing.

After CMP, the normalized CL spectrum exhibits a relative enhancement of near-band-edge emissions ($DG_1$–$DG_2$) and a noticeable suppression of deep-level transitions ($DT_1$–$DT_3$), reflecting improved optical quality. This spectral redistribution indicates that CMP effectively removes non-stoichiometric surface layers and defect clusters, thereby reducing the density of luminescent trap states. The corresponding SEM (Fig. 3b) and CL mapping (Fig. 3c) images further support this interpretation, showing a locally smooth surface with spatially uniform emission intensity.

Figure 4 summarizes the thermal conductivity measurements of the polished rutile $GeO_2$ films performed by TDTR. The normalized temperature decay signal (termed as "ratio" here) obtained using experiments (symbols) as a function of delay time (Fig. 4a) was fitted using a multilayer heat-diffusion framework that incorporated the known thermal properties of the Al transducer layer and the r-$TiO_2$ substrate (dashed line). The experimental data show an excellent fit with the simulated curves, validating the thermal model. Measurements were conducted with multiple scans at multiple points across the film surface to ensure reproducibility. As shown in Fig. 4b, the extracted cross-plane thermal conductivity values are consistent across all locations, yielding an average of $52.9 \pm 6.6$ W m$^{-1}$ K$^{-1}$ at 300 K. For comparison, standard sapphire ($\alpha$-$Al_2O_3$) and r-$TiO_2$ reference samples measured under identical conditions yielded thermal conductivities of 34.8 W m$^{-1}$ K$^{-1}$ and 10.8 W m$^{-1}$ K$^{-1}$, respectively. The significantly higher thermal conductivity r-$GeO_2$ highlights its superior phonon transport relative to these benchmark oxides. The thermal conductivity value also surpasses that of $\beta$-$Ga_2O_3$ (~27 W m$^{-1}$ K$^{-1}$ along *b*-axis) and is comparable to or greater than many III–nitride systems such as AlGaN. The observed high thermal conductivity can be attributed to the tetragonal rutile lattice symmetry and strong Ge–O bonding features, which promote elevated acoustic phonon velocities and reduce phonon-scattering channels. The surface planarization achieved via CMP also plays a crucial role. The removal of high-aspect-ratio grains and defect clusters reduces phonon-boundary scattering, allowing more efficient heat conduction across the film. The measured thermal conductivity primarily reflects the intrinsic heat transport within the r-GeO$_2$ layer, as the experimental configuration was designed to minimize substrate influence and exclude interfacial thermal resistance effects.

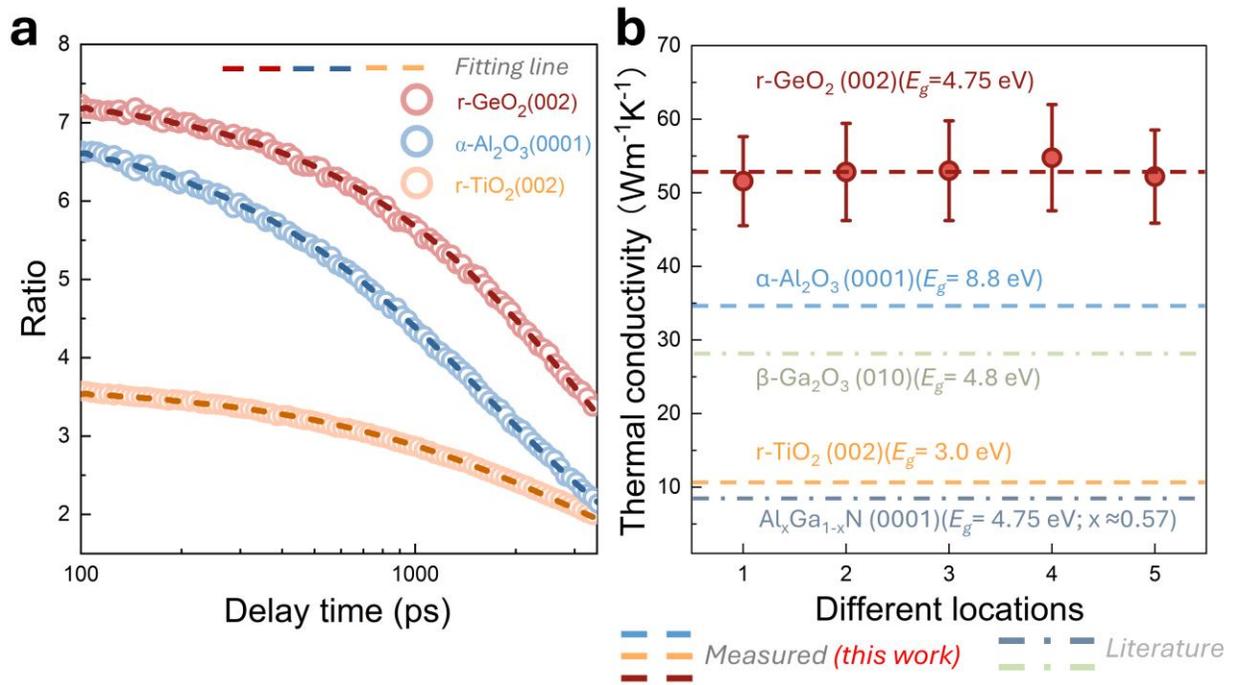

**FIG. 4.** (a) TDTR ratio versus delay time for r-GeO$_2$ (002), α-Al$_2$O$_3$ (0001), and r-TiO$_2$ (002) with best-fit curves obtained using a multilayer thermal model. (b) Extracted thermal conductivity of r-GeO$_2$ (002) across five surface locations, yielding an average value of 52.9 ± 6.6 W m$^{-1}$ K$^{-1}$. For comparison, α-Al$_2$O$_3$ (0001) (34.8 W m$^{-1}$ K$^{-1}$, $E_g$=8.8 eV[45]), r-TiO$_2$ (002) (10.8 W m$^{-1}$ K$^{-1}$, $E_g$ = 3.0 eV[46,47]), β-Ga$_2$O$_3$ (010) (27 W m$^{-1}$ K$^{-1}$,[48] $E_g$=4.8 eV[49,50]), and Al$_{0.57}$Ga$_{0.43}$N (0001) (8.8 W m$^{-1}$ K$^{-1}$,[51] $E_g$=4.75 eV[52]) are shown for reference, highlighting the competitive thermal and optical properties of r-GeO$_2$ among UWBG oxides.

In summary, r-GeO$_2$ thin films were grown by MOCVD using the SDSC approach and subsequently polished by CMP to achieve an atomically smooth surface (RMS ≈ 1 Å). The films exhibited single-phase r-GeO$_2$ (002) orientation with fourfold epitaxy on r-TiO$_2$ (001), and both as-grown and CMP-treated samples showed similar rocking-curve FWHM (~0.28°), confirming preserved crystalline quality after polishing. CL analysis showed the suppression of defect-related deep-level emissions and the dominance of intrinsic blue–green bands at 2.64 eV and 2.38 eV after polishing. The TDTR-measured thermal conductivity reached 52.9 ± 6.6 W m$^{-1}$ K$^{-1}$, surpassing sapphire (~34.8 W m$^{-1}$ K$^{-1}$) and rutile TiO$_2$ (~10.8 W m$^{-1}$ K$^{-1}$) and closely matching theoretical predictions (~37–58 W m$^{-1}$ K$^{-1}$). These results demonstrate that combining SDSC growth with CMP planarization enables the structurally coherent r-GeO$_2$ films with improved phonon transport, underscoring their potential for thermally robust UWBG power devices.

## AUTHOR DECLARATIONS

### Conflict of Interest

The authors have no conflicts to disclose.

### Author Contributions

**Imteaz Rahaman**: Data curation (equal); Formal analysis (equal); Investigation (lead); Methodology (equal), Writing-original draft (lead). **Michael E. Liao**: Data curation (equal); Formal analysis (equal); Methodology (equal); Writing – review & editing (supporting). **Ziqi Wang**: Data curation (equal); Formal analysis (equal); Methodology (equal); Writing – review & editing (supporting). **Eugene Y. Kwon**: Data curation (supporting); Methodology (supporting). **Rui Sun**: Data curation (supporting); Methodology (supporting). **Botong Li**: Writing – review & editing (supporting). **Hunter D. Ellis**: Writing – review & editing (supporting). **Bobby Duersch**:

## ACKNOWLEDGEMENT

The authors acknowledge the support from the University of Utah start-up fund and PIVOT Energy Accelerator Grant No. U-7352FuEnergyAccelerator2023. Z.W. and J.L. acknowledge financial support from the National Science Foundation under the award number CBET-1943813 for the TDTR measurements. R.S. and D.S. acknowledge financial support from the Department of Energy under the award number DE-SC0020992 for the metal transducer deposition. This work made use of the Nanofab EMSAL shared facilities of the Micron Technology Foundation Inc. Microscopy Suite, sponsored by the John and Marcia Price College of Engineering, Health Sciences Center, Office of the Vice President for Research. In addition, it utilized the University of Utah Nanofab shared facilities, which are supported in part by the MRSEC Program of the NSF under Award No. DMR-112125.

## DATA AVAILABILITY

The data that supports the findings of this study are available from the corresponding authors upon reasonable request.